\documentclass[aps,prl,twocolumn,showpacs]{revtex4}
\usepackage{graphicx}
\usepackage{multirow}
\usepackage{rotating}
\begin{document}
\title{{\it Ab initio} charge, spin and orbital energy scales in LaMnO$_{3}$}
\author{R. Tyer$^{1,2}$, W.M. Temmerman$^1$, Z. Szotek$^1$, G. Banach$^1$, 
A. Svane$^{3}$, L. Petit$^{3}$, G.A. Gehring$^2$}
\affiliation{$^1$ Daresbury Laboratory, Daresbury, Warrington WA4 4AD, UK}
\affiliation{$^2$ Department of Physics and Astronomy, University of Sheffield, Sheffield, S3 7RH, UK}
\affiliation{$^3$ Department of Physics and Astronomy, University of Aarhus, DK-8000 Aarhus, Denmark}
\date{\today}
% ************************ abstract *********************************
\begin{abstract}
The first-principles SIC-LSD theory is utilized to study electronic, magnetic and orbital
phenomena in LaMnO$_{3}$. The correct ground state
is found, which is antiferro orbitally ordered with the spin magnetic moments 
antiferromagnetically aligned. 
%Minimization of the hybridisation between the localised 
%e$_g$ states and delocalized p orbitals of the adjacent oxygens provides the {\it ab initio} 
%understanding of the underlying physics. 
Jahn-Teller energies are found to be the largest energy scale.
In addition it is the Jahn-Teller interaction which is the dominant effect in realising  orbital order,
and the electronic effects alone do not suffice.
%The study correctly describes the pressure
%induced delocalization of an e$_{g}$ electron in LaMnO$_{3}$. 
\end{abstract}
\maketitle
%\nopagebreak
%\begin{multicols}{2}
%%%%%%%%%%%%%%%%%%%%%%%%%%%%%%%%%%%%%%%%%%%%%%%%%%%%%%%%%%%%%%%%%%%%%%%%%%%%%%%%%%%%%%%%%%%

\begin{figure}
%\begin{figure}[htbp]
%\begin{center}
\includegraphics[scale=.37]{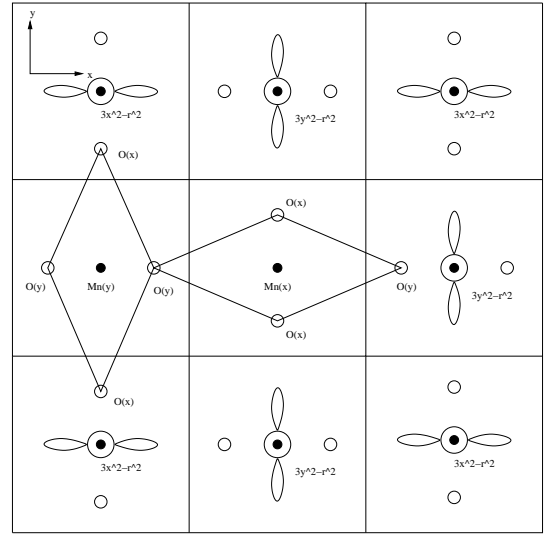}
\caption{Simplified view of the manganese oxygen plane in LaMnO$_3$. The small black circles represent 
the manganese atoms, while the open circles mark the positions of the oxygens. The distortion has been 
exaggerated to show clearly that the manganese oxygen bonds alternate between long and short in the x 
and y directions (designated Mn(x) and Mn(y) respectively), which is associated with the antiferro OO 
%orbital ordering 
of the $d_{3x^2-r^2}$ and $d_{3y^2-r^2}$ orbitals.}
\label{OOfig}
%\end{center}
\end{figure}

There are several transition metal compounds in which the orbital degeneracy is broken spontaneously.
Examples are KCuF$_3$,\cite{KCuF3} V$_2$O$_3$,\cite{V2O3} 
and the manganites,\cite{coey} which are the 
subject of the present study. 
In the manganites, the crystal field associated with MnO$_6$ octahedra
splits the manganese $d$ levels into a lower lying t$_{2g}$ triplet and an upper e$_{g}$ doublet.
The t$_{2g}$ states are highly localized whereas an electron in one of 
the e$_{g}$ states is potentially
itinerant. The Mn$^{4+}$ ion in CaMnO$_3$  has a fully occupied 
majority t$_{2g}$ manifold and empty e$_g$ states,
which form a strongly localized core spin S=$\frac{3}{2}$.
The Mn$^{3+}$ ion in LaMnO$_3$, on the other hand, has an additional $d$ 
electron which, due to the strong 
intra-atomic exchange, populates one of the e$_{g}$ states, 
forming an $S=2$ spin, and which gives rise to 
a Jahn-Teller (JT) instability. 
In this system each of the oxygen ions is a
neighbour to two Mn ions and hence the local distortions of the lattice must
be arranged in such a way as to minimize the energy for the whole crystal.
%In this material 
This gives rise to Mn-O bond lengths of 1.90 and 2.18\AA\,
within the manganese oxygen plane, compared with 1.96\AA\, for the hypothetical cubic system.
The e$_g$ states rotate to form an orbitally ordered lattice of $d_{3x^2-r^2}$ and $d_{3y^2-r^2}$ orbitals 
in the manganese oxygen plane\cite{Goodenough}, shown schematically in Fig. \ref{OOfig}.
%Doping LaMnO$_3$ with Ca, to form the manganite series $La_{1-x}Ca_xMnO_3$,
%is equivalent to doping the e$_{g}$ band with holes.
%It is this hole doping in conjunction with the strong coupling between the spin, charge and orbital 
%degrees of freedom which gives rise to the rich phase diagram exhibited by the manganites \cite{vanderBrink}. 

%%%%%%%%%%%%%%%%%%%%%%%%%%%%%%%%%%%%%%%%%%%%%%%%%%%%%%%%%%%%%%%%%%%%%%%%%%%%%%%%%%%%%%%%%%%
%%%%%%%%%%%%%%%%%%%%%%%%%%%%%%%%%%%%%%%%%%%%%%%%%%%%%%%%%%%%%%%%%%%%%%%%%%%%%%%%%

%%%%%%%%%%%%%%%%%%%%%%%%%%%%%%%%%%%%%%%%%%%%%%%%%%%%%%%%%%%%%%%%%%%%%%%%%%%%%%%%%
%This paper reports a first-principles study of these intricate phenomena, with the emphasis on orbital
This paper reports a first-principles study of charge, 
spin and orbital energy scales in LaMnO$_{3}$, with the emphasis on orbital
ordering (OO), based upon the self-interaction corrected (SIC) local spin-density (LSD) 
theory,\cite{SIC,Brisbane} which allows $d$-electron localization to be distinguished from 
itinerancy, in an {\it ab initio} manner.\cite{Nature,PRL} 
%%%%%%%%%%%%%%%%%%%%%%%%%%%%%%%%%%%%%%%%%%%%%%%%%%%%%%%%%%%%%%%%%%%%%%%%%%%%%%%%%%%%%%%%%%%
%  1. Removed to make space for stuff suggested by Gillian                                %
%%%%%%%%%%%%%%%%%%%%%%%%%%%%%%%%%%%%%%%%%%%%%%%%%%%%%%%%%%%%%%%%%%%%%%%%%%%%%%%%%%%%%%%%%%%
%and has been successfully applied to describe the valency 
%of rare earths \cite{Nature} and the metal-insulator transition in YBCO \cite{PRL}.
%The SIC-LSD method can determine if a Mn $d$-electron is localized or
%itinerant by comparing the total energies of calculations treating the $d$-electron as either 
%localized, i.e. moving in the SIC potential, or delocalized, namely responding to the effective 
%LSD potential. The resulting energy difference is used to establish whether the energy 
%gain upon the $d$-electron localization is favoured over band formation and hybridisation.
%%%%%%%%%%%%%%%%%%%%%%%%%%%%%%%%%%%%%%%%%%%%%%%%%%%%%%%%%%%%%%%%%%%%%%%%%%%%%%%%%%%%%%%%%%%
Not only does the SIC allow the localization of an orbital of any symmetry, but the SIC-LSD total energies
can be minimized with respect to the number of localized orbitals and their orientation.
In addition, minimization with respect to the number of localized orbitals yields valency which
is defined as an integer number of electrons available for band formation, 
$N_{val} = Z - N_{core} - N_{SIC}$, where Z is the atomic number, N$_{core}$ is the number of core 
(and semi-core) states, and N$_{SIC}$ the number of self-interaction corrected (localized) states.
%Thus, the SIC-LSD theory provides a novel method of performing {\it ab initio} investigations of 
%OO, aiming to identify the ground state configuration of the spin and orbital ordering.
%Furthermore, in the SIC-LSD a single orbital is chosen. 
%It is localised but its occupation is determined self consistently. 
%This automatically breaks the $e_{g}$ orbital symmetry and 
%gives a nonzero value for the local orbital order.  
Furthermore, in the SIC-LSD the occupation of each localized orbital 
is determined self-consistently. When localizing an $e_{g}$ orbital
its symmetry is  automatically broken leading to a nonzero value for the local orbital order.
For example if the orbital that is localized corresponds to $d_{3z^{2}-r^{2}}$ then the occupation 
of this orbital and its partner, $d_{x^{2}-y^{2}}$, will be different. This difference, because
of hybridization effects, will not be equal 
to unity as it would be in a fully localized picture. 
The SIC-LSD provides the quantitative comparison of all energies for LaMnO$_3$.  
This material has a distorted cubic structure with OO as shown in  Fig. \ref{OOfig} 
and is an A-type antiferromagnet (A-AFM) such that the moments in the x-y planes shown in  Fig. \ref{OOfig} 
are ordered ferromagnetically and the planes are stacked antiferromagnetically along the c axis.  
Although the dominant distortion about a Mn site is tetragonal there are other components so that 
the net symmetry is lower. 
%%%%%%%%%%%%%%%%%%%%%%%%%%%%%%%%%%%%%%%%%%%%%%%%%%%%%%%%%%%%%%%%%%%%%%%%%%%%%%%%%%%%%%%%%%%
We have established that the relevant energies are:
%, in descending order:
change in the total energy due to the distortion, localization energy, 
the purely electronic contribution to the 
orbital ordering energy  (i.e., 
the change in the electronic energy induced by OO in the absence of distortion
%effects on the band electrons induced by the OO
), and the magnetic ordering energy;
%from purely electronic effects; 
these energy scales are summarized in Table \ref{summary}.  
%%%%%%%%%%%%%%%%%%%%%%%%%%%%%%%%%%%%%%%%%%%%%%%%%%%%%%%%%%%%%%%%%%%%%%%%%%%%%%%%%%%%%%%%%%%
\begin{table}
%\begin{table}[htbp]
\caption{Summary of all energy scales in LaMnO$_{3}$. }
%\begin{center}
\begin{tabular}{|l|c|}
\hline
Quantity & Energy gain  \\
& (mRy per formula unit) \\
\hline
Electronic energy gain  &  \\
due to the distortion & $\sim 40$ \\
\hline
Localisation energy & $\sim 20$ \\
\hline
Purely electronic contribution to & \\
the orbital ordering energy & $\sim 5$ \\
\hline
Relative stability of  & \\ 
the A type antiferromagnetic & \\ 
state compared with ferromagnetic & $\sim 1$ \\
\hline
\end{tabular}
\label{summary}
%\end{center}
\end{table}
%%%%%%%%%%%%%%%%%%%%%%%%%%%%%%%%%%%%%%%%%%%%%%%%%%%%%%%%%%%%%%%%%%%%%%%%%%%%%%%%%
%%%%%%%%%%%%%%%%%%%%%%%%%%%%%%%%%%%%%%%%%%%%%%%%%%%%%%%%%%%%%%%%%%%%%%%%%%%%%%%%%%%%%%%%%%%

The largest electronic energy %, of about 40 mRy per formula unit, 
is that due to the distortion
i.e., the gain in energy by imposing OO in the distorted crystal structure.
The second largest energy in the problem is the localization energy which is the difference 
between the energies of 
localized and delocalized $e_{g}$ states.
The value is about 20 mRy per formula unit, depending slightly 
on the state of magnetisation and distortion.
%[{\bf Axel:} What about comparing localized and delocalized e$_g$, i.e., Mn(3+) and Mn(4+)? ]
The magnetisation comes last, as the smallest energy scale,  
behind the purely electronic contribution to the orbital ordering energy.
The oxygen displacements around site Mn(x) are such that the site symmetry  has a large,  
$2{\epsilon}_{xx}-{\epsilon}_{yy}-{\epsilon}_{zz}$, distortion, as well as 
other smaller distortions that break the symmetry between y and z. 
The Mn(y) site has a similar distortion.  
Table \ref{OOmaster} contains all SIC-LSD energies 
when a given e$_{g}$ orbital has been localized 
on a given site, for A-AFM spin ordering. 
All calculations were done for the experimental distorted structure.\cite{Elemans}
%%%%%%%%%%%%%%%%%%%%%%%%%%%%%%%%%%%%%%%%%%%%%%%%%%%%%%%%%%%%%%%%%%%%%%%%%%%%%%%%%%%%%%%%%%%
%%%%%%%%%%%%%%%%%%%%%%%%%%%%%%%%%%%%%%%%%%%%%%%%%%%%%%%%%%%%%%%%%%%%%%%%%%%%%%%%%
\begin{table}[htbp]
\caption{Results for several scenarios where the three t$_{2g}$ orbitals and an additonal e$_g$ state 
are localized on the manganese 
atoms in fully distorted A-AFM structure of LaMnO$_3$. 
The first column gives the numbering of the scenarios which involve AF OO and ferro (F) OO. 
Columns 2 and 3 indicate which e$_g$ orbital is localized on the manganese atoms with the long 
manganese-oxygen bonds in the x and y directions respectively.
The relative energies $\Delta$E, with respect to the ground state 
with the localized $d_{3x^2-r^2}$ orbital on the 
Mn(x) sites and the $d_{3y^2-r^2}$ state on the Mn(y) atoms, 
is displayed in column 4 (A-AFM). Column
5 refers to relative energies $\Delta E_{d}$ obtained 
from a localized model which incorporates 
only the orbital ordering in a purely tetragonal distortion.(see text)   
%$\Delta$$E_{SIC}$ is the relative energy loss, with respect to the ground state, of the SIC states (see text).
}
\begin{center}
\begin{tabular}{c||c|c|c|c|c|}
  & & & &\multicolumn{2} {c|} {A-AFM} \\
\hline
 & \#& Mn(x) & Mn(y)  &  $\Delta$E (mRy)  & $\Delta E_{d} (mRy)$ \\
% &       &        &  (eV) &  (mRy)    &  (mRy)   \\
\hline
\multirow{4}{0.1in}{\begin{sideways}{AF OO}\end{sideways}} & 1& $d_{3x^2-r^2}$ & $d_{3y^2-r^2}$  &   0  & 0 \\
 &2 & $d_{z^2-x^2}$  & $d_{y^2-z^2}$   & 8.5 & 20 \\
 &3 & $d_{3y^2-r^2}$ & $d_{3x^2-r^2}$  & 65.2 & 59 \\
 &4 & $d_{y^2-z^2}$  & $d_{z^2-x^2}$   & 78.1 & 78 \\
\hline
\multirow{6}{0.05in}{\begin{sideways}{F OO}\end{sideways}} &5 & $d_{x^2-y^2}$  & $d_{x^2-y^2}$ & 27.2 & 20 \\
  &6 & $d_{3x^2-r^2}$ & $d_{3x^2-r^2}$ & 39.7 & 29 \\
  &7 & $d_{3y^2-r^2}$ & $d_{3y^2-r^2}$  & 40.5 & 29 \\
  &8 & $d_{3z^2-r^2}$ & $d_{3z^2-r^2}$  & 49.5 & 59 \\
  &9 & $d_{z^2-x^2}$  & $d_{z^2-x^2}$   & 56.6 & 49 \\
  &10 & $d_{y^2-z^2}$  & $d_{y^2-z^2}$   & 57.3 & 49 \\
%\hline
\end{tabular}
\label{OOmaster}
\end{center}
\end{table}
%%%%%%%%%%%%%%%%%%%%%%%%%%%%%%%%%%%%%%%%%%%%%%%%%%%%%%%%%%%%%%%%%%%%%%%%%%%%%%%%%
%%%%%%%%%%%%%%%%%%%%%%%%%%%%%%%%%%%%%%%%%%%%%%%%%%%%%%%%%%%%%%%%%%%%%%%%%%%%%%%%%
%%%%%%%%%%%%%%%%%%%%%%%%%%%%%%%%%%%%%%%%%%%%%%%%%%%%%%%%%%%%%%%%%%%%%%%%%%%%%%%%%

The following points may be seen from this table.
First, the most favourable state is the antiferro orbital ordering (AF OO) on the correct site 
coupled with antiferromagnetic spin ordering as observed experimentally.
Second, the results for scenarios 6 and 7 should be identical from symmetry. 
They were calculated independently and the difference between them represents an estimate of the 
error on our calculations; the same is true for scenarios 9 and 10. 
%We have confidence in our results to the nearest mRy. 
Third, the size of the total electronic energy favouring the orbital ordering is found from 
the most and least favorable AF scenarios 1 and 4, which differ 
 by $\sim ~78$ mRy, half of which is then the energy gain due to OO, i.e. the largest energy
scale of Table \ref{summary}. 
More explicitly, the states in scenarios 1 and 3 have identical pattern of antiferro orbits. 
The difference between them is that in the former case the lobes of the orbits match the lattice 
distortion, while in the latter
case the orbit is misaligned with respect to the distortion.  
Band structure effects favour these states equally and the difference in their energies 
represents the importance of the lattice distortion for the OO energy.
%Fourth, the large orbital ordering energy only exists in the distorted phase, being
%reduced by approximately a factor of 20  in the cubic phase.

The energy scale that characterises the stability of A-AFM versus FM
is the smallest energy scale, at $\sim 1$ mRy, which is similar to energies that 
gave correct exchange constants for NiO(100) surface.\cite{Diemo}
%, leading to
%exchange constants in agreement with experiment
%Therefore
%may be estimated from the difference in energy between the A-AFM and FM 
%state compared with ferromagnetic & $\sim 1$ \\
%ground states with the most favourable orbitals, and equals 15 mRy. 
%Mean field theory for S=2 predicts that the transition temperature is related to the energy gain 
%per site, $u$, by $k_{B}T_{c}=u$. This gives an estimate for $T_{c}$ = 165 K which is rather higher 
%than the experimental value of 120 K as expected for mean field theory.
Thus, the ordering sequence of the rows will be the same for the A-AFM and the FM structures,
which indicates that the dominant cause of the orbital ordering is not the A-AFM magnetic 
structure, as has been postulated.\cite{Medvedeva,Okamoto} 
%%%%%%%%%%%%%%%%%%%%%%%%%%%%%%%%%%%%%%%%%%%%%%%%%%%%%%%%%%%%%%%%%%%%%%%%%%%%%%%%%%%%%%%%%%%

It is instructive to see how far a localized model that assumes only a locally 
pure tetragonal distortion can account for the {\it ab initio} results.  
The energy per site is given by $u=u_{0}-u_{D}cos2\theta$ where $u_{0}$ and $u_{D}$ 
are the energies that do not and do depend on the distortion respectively.  
The angle $\theta$ defines the orbit.  For the site Mn(x) we have $\theta=0:d_{3x^{2}-r^{2}}$, 
$\theta=\pm{\pi/6}:d_{x^{2}-y^{2}},d_{x^{2}-z^{2}}$, $\theta=\pm{2\pi/3}:d_{3y^{2}-r^{2}},d_{3z^{2}-r^{2}}$,
$\theta={\pi/2}:d_{y^{2}-z^{2}}$, and equivalently for the Mn(y) site.  
The energies of scenarios 1 and 4 may be used to fix $u_{0}=u_{D}= 39$  mRy for the A-AFM  phase.
% and $u_{0}= 49$ mRy, $u_{D}= 34$ mRy for the FM phase.  
The results given in column 5 
of Table \ref{OOmaster} are found 
using this expression. 
It is seen that the order of energies obtained by this simple model 
reproduces the trends seen in the {\it ab initio} results.  
%of the energy levels is correct and 
Moreover, most {\it ab initio} energies lie within 10 mRy of this very simple model.
It appears that the discrepancy between the model and the first principles results is 
mostly due to other distortions than tetragonal. 
In order to investigate this we have performed {\it ab initio} calculations  
for a crystal with a pure tetragonal distortion (of the observed magnitude).  
In this case we found a slightly increased  value for $u_{D}$ ( 42 mRy) 
and the deviations between the model and 
the first principles calculation are reduced to 5 mRy, for all scenarios.
The size of these remnant  energy fluctuations designates the scale 
of the band electrons' contribution to the total energy. The imposed OO, lattice
distortion and magnetic structure induce changes to the conduction electron states,
which leads to changes of the order of 5 mRy to the total energy.
This represents the third energy scale of Table \ref{summary}, which is 
%significantly smaller than any of the other scales considered.
the second smallest of the energy scales considered.
%The difference between the values of $u_{D}$ for the two magnetic phases 
%also gives a measure for the size of the band-electronic contribution, and still
Another estimate for the band-electronic effects
was obtained by evaluating the OO total energy variations in the cubic phase.  
Our calculations gave 3.7 and 8.3 mRy in this case, for FM and A-AFM ordering respectively, 
i.e. OO is drastically suppressed without the lattice distortion.  
% The following I don't think we can use here, since we have already accounted for this energy
% variation with the theta-model. Axel :
% A further indication that the electronic contribution is small came from comparing 
% the energies of scenarios 1 and 2.  The orbital ordering is the same in both cases but 
% the difference is that in scenario 2 it is in the incorrect position with respect to the lattice.  
%We find that the simple model gives a result here that is too low and that this cannot be accounted 
%for by electronic orbital effects as this would reduce the contribution to $u_{D}$ above 
%and hence make the agreement with the model worse.
%We believe that the other distortions are contributing to the discrepancy between the model and 
%the first principles results.  
%%%%%%%%%%%%%%%%%%%%%%%%%%%%%%%%%%%%%%%%%%%%%%%%%%%%%%%%%%%%%%%%%%%%%%%%%%%%%%%%%%%%%%%%%%%

To proceed with the analysis of the energetics of OO in the manganites,
we now consider the lattice distortion mode.
The strength of the orbital-lattice interaction may be deduced
from the measured values of the
elastic constants and the size of the lattice distortion.  
%{\bf Axel: I think we may skip index $\alpha$}
To see this, we follow Ref. \onlinecite{Millis} and
 consider an expansion of the total energy of an electronic system at T =0 
in terms of a local distortion in the unit cell at ${\bf R}_{i}$, ${\epsilon}_{i}$.% ^\alpha$,  
%with $\alpha$ labeling the nonzero component of the order parameter.
The restoring force for that mode is characterized by the force constant $K$,
and the elastic energy of the distortion  has a simple harmonic dependence on amplitude.
Countering this is the nonzero orbital order, which contributes a negative energy:
\begin{eqnarray}
%U_{tot}& = & U_{0}+U^{\alpha}_{el}(\{{\epsilon}_{i}^\alpha\})+
%\frac{1}{2}\sum_{i}K({\epsilon}_{i}^\alpha)^{2} \nonumber \\
%       & =& U_{0}+U^{\alpha}_{el}(0)-\sum_{i}{{u}_{i}^\alpha}{{\epsilon}_{i}^\alpha}+
%    \frac{1}{2}\sum_{i}K({\epsilon}_{i}^\alpha)^{2}. \label{eq:one}
U_{tot}& = & U_{0}+U_{el}(\{{\epsilon}_{i}\})+
\frac{1}{2}\sum_{i}K({\epsilon}_{i})^{2} \nonumber \\
       & =& U_{0}+U_{el}(0)-\sum_{i}{{u}_{i}}{{\epsilon}_{i}}+
    \frac{1}{2}\sum_{i}K({\epsilon}_{i})^{2}. \label{eq:one}
\end{eqnarray}
In deriving this expression we assumed that only linear terms in ${\epsilon}_{i}$  
were required for the OO energy, and defined 
%${{u}_{i}}$ by 
%${{u}_{i}^\alpha}$ by 
%\begin{equation}
\begin{displaymath}
%{{u}_{i}^\alpha}=
%-\frac{\partial U_{el}^{\alpha}(\{{\epsilon}_{i}^{\alpha}\})}{\partial \epsilon_{i}^{\alpha}}.  
{{u}_{i}}=
-\frac{\partial U_{el}(\{{\epsilon}_{i}\})}{\partial \epsilon_{i}}.  
\end{displaymath}
%\end{equation}
%The second term in Eq. (1), 
%%$U^{\alpha}_{el}(0)$, represents the small band-electronic contribution 
%$U_{el}(0)$, represents the small band-electronic contribution 
%to the OO energy, while the third term we identify with $u_D$.  
At equilibrium, the elastic and OO  energies must balance, and the distortion parameter is 
found by minimising the energy, i.e., % with respect to $\epsilon_{i}^\alpha$, i.e.,
$\epsilon_{i}=\frac{u_{i}}{K}$.
%$\epsilon_{i}^\alpha=\frac{u_{i}^\alpha}{K}$.
Therefore the total lowering of energy due 
to the lattice distortion and 
OO is found by substituting 
the above expression for 
%$\epsilon_{i}^\alpha$ into equation (\ref{eq:one}):
$\epsilon_{i}$ into equation (\ref{eq:one}):
\begin{displaymath}
%\Delta U^{\alpha} =  U^{\alpha}_{el}(0)-\frac{1}{2}\sum_{i}K({\epsilon}_{i}^\alpha)^{2} 
%        = U^{\alpha}_{el}(0)-\frac{1}{2}\sum_{i} {u}_{i}^\alpha {\epsilon}_{i}^\alpha . 
\Delta U =  U_{el}(0)-\frac{1}{2}\sum_{i}K({\epsilon}_{i})^{2} 
        = U_{el}(0)-\frac{1}{2}\sum_{i} {u}_{i} {\epsilon}_{i} . 
\end{displaymath}
The distortions are known in the manganites, $\epsilon_i\sim 0.30$ \AA,\cite{Elemans} 
as are also the optic phonon frequencies, leading to $K\sim 12.5$ eV/\AA$^2$.\cite{Millis}  
%{\bf Axel: are these the  correct numbers? - energy per formula unit?}
Hence, we can determine an experimental value of 
%${u}_{i}^\alpha$ 
${u}_{i} = 3.5$ eV/\AA. The SIC-LSD theory leads to the estimate 
${u}_{i} \sim u_D/\epsilon_i \sim 1.8 $ eV/\AA, which is to be considered in excellent
agreement, given the general
uncertainty, not least in estimating the appropriate value of $K$.\cite{Millis} 
%The SIC-LSD theory leads to the estimate 
%, and from this the electronic OO energy $u_D\sim 90$ mRy. 
%{\bf Axel: Which is a factor of 2 too big - have I made a mistake?}
%Thus we see that provided the distortions are measured,
% the total energy due to the interaction 
%with the lattice is known,\cite{Millis} and the Jahn Teller ordering 
%energy of $\sim$ 0.6 eV=44 mRy is obtained which is close to the value of the 
%orbital energy due to the distortion 
%from this first principles calculation as given by $u_{D}$. 
Otherwise stated, neglecting $U_{el}(0)$, a total Jahn-Teller ordering 
energy of  $\sim \frac{1}{2}u_D \sim 20$ mRy is arrived at, which is in rather good agreement with the estimate of 
%$\sim 0.48 to 0.58$ eV $=35 to 43$ mRy of Ref. \onlinecite{Millis}.
0.48 to 0.58 eV =35 to 43 mRy of Ref. \onlinecite{Millis}.
\begin{figure}
%\begin{figure}[tbp]
%\begin{figure}[htp]
%\begin{center}
%\includegraphics[scale=.37]{fav_new.eps}
\includegraphics[bb= 0 0 588 740,viewport=0 100 588 740,scale=.37]{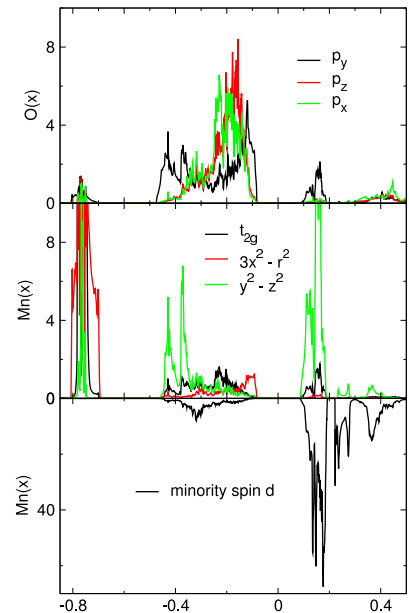}
\caption {The orbital and spin 
resolved densities of states 
corresponding to 
scenario 1 (see Table \ref{OOmaster}),
%the localized $d_{3x^2-r^2}$ orbital on Mn(x) 
%sites and the $d_{3y^2-r^2}$ on the Mn(y) sites, 
i.e. with the e$_g$ lobes oriented along the long Mn-O bonds. 
The top panel shows the majority spin partial p densities of states for the O(x) sites, while the 
middle and lower panels show the  majority and minority (summed over all d orbitals) 
spin partial d densities of states for the Mn(x) site respectively. 
The energies (in Ry) are relative to the Fermi energy at $E=0$. 
The partial densities of states are in units of states/spin/Ry.
}
\label{fig:fav}
%\end{center}
\end{figure}
%%%%%%%%%%%%%%%%%%%%%%%%%%%%%%%%%%%%%%%%%%%%%%%%%%%%%%%%%%%%%%%%%%%%%%%%%%%%%%%%%
To shed more light on the hybridization effects involved in OO, 
we inspect in detail the magnetic $m$ and orbital $l$ quantum numbers 
resolved densities of states for two OO configurations, namely
scenarios 1 and 3 of Table \ref{OOmaster} which are shown in 
Figs. \ref{fig:fav} and \ref{fig:unfav} respectively. 
%The former figure refers to the case where the lobe-like e$_g$ orbitals 
%are oriented along the long Mn-O bonds (see Fig. \ref{OOfig}),
%which is the ground state configuration and will be referred to 
%as "favourable", while the latter figure corresponds to the situation where the 
%lobe-like e$_g$ orbitals are oriented along the short Mn-O bonds referred to as "unfavourable".
%Specifically, Figs. \ref{fig:fav} and \ref{fig:unfav} refer to the scenarios 
%1 and 3 of Table \ref{OOmaster} respectively. 
%Specifically, Fig. \ref{fig:fav} refers to the DOS where the localised orbitals 
%are the t$_{2g}$'s and the $d_{3x^2-r^2}$ on the Mn(x) site, whilst in Fig. \ref{fig:unfav}
%the localised orbitals on this Mn site are t$_{2g}$'s and $d_{3y^2-r^2}$. 
Localised orbitals 
are seen to have only marginal admixture (below 0.1 electron) of the delocalized $e_{g}$ state. The latter   
(in green) can be found both in the valence and conduction bands.
The localized orbitals do still
contribute some weight in the valence and conduction bands of the Mn(x) majority DOS in Figs. \ref{fig:fav}
and \ref{fig:unfav}. In particular, t$_{2g}$ states contribute to the valence and conduction bands.
This reflects the slight
hybridization of the t$_{2g}$ states, showing that they are not of pure ionic character but acquire
some covalency. For the localized e$_g$ state this hybridization increases and is most 
pronounced for $d_{3y^2-r^2}$ in Fig. \ref{fig:unfav}. Here we find $d_{3y^2-r^2}$ states in both the conduction
band and at the top of the valence band, summing up to 0.2 electrons. 
Obviously, this localized state, whose lobes point along the short
Mn-O bond length, hybridizes strongly with the $p_x$ orbital of O(y) site   
and this can be seen as the $p_x$  weight in the conduction band of O(y) partial DOS in Fig. \ref{fig:unfav}.
In contrast the $d_{3x^2-r^2}$ weight in the valence and conduction bands for scenario 1 
is less than the $d_{3y^2-r^2}$ weight in scenario 3.
In scenario 1 the e$_g$ orbital is more localized: 0.82 localized $d_{3x^2-r^2}$ electrons 
versus 0.74 $d_{3y^2-r^2}$ in the DOS peak around -0.8 Ry. 
%In the favourable scenario  better localization of the e$_g$ orbital is obtained because the $d_{3x^2-r^2}$
%orbital points along the long Mn-O bond leading also to reduced hybridization with the $p_x$ on the
%O(y) site. This can be seen in the reduced O $p$ weight in the conduction bands.

%%%%%%%%%%%%%%%%%%%%%%%%%%%%%%%%%%%%%%%%%%%%%%%%%%%%%%%%%%%%%%%%%%%%%%%%%%%%%%%%%%%% 

%%%%%%%%%%%%%%%%%%%%%%%%%%%%%%%%%%%%%%%%%%%%%%%%%%%%%%%%%%%%%%%%%%%%%%%%%%%%%%%%%
%%%%%%%%%%%%%%%%%%%%%%%%%%%%%%%%%%%%%%%%%%%%%%%%%%%%%%%%%%%%%%%%%%%%%%%%%%%%%%%%%
%%%%%%%%%%%%%%%%%%%%%%%%%%%%%%%%%%%%%%%%%%%%%%%%%%%%%%%%%%%%%%%%%%%%%%%%%%%%%%%%%
\begin{figure}
%\begin{figure}[hbp]
%\begin{center}
\includegraphics[scale=.37]{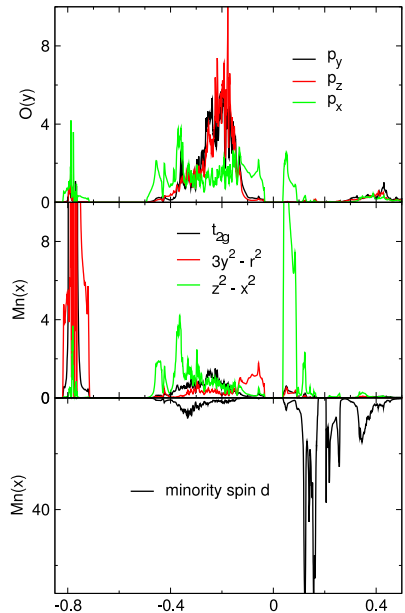}
\caption{The same as Fig. \ref{fig:fav} but for 
scenario 3 (see Table \ref{OOmaster}),
%the case where the $d_{3y^2-r^2}$ orbital is localised on Mn(x) 
%sites, and the $d_{3x^2-r^2}$ on the Mn(y) sites, 
i.e., the e$_g$ lobes are oriented along the short Mn-O 
bonds. The top panel shows the majority spin partial p densities of states for the O(y) sites. 
%while the 
%middle and lower panels show the majority and minority (summed over all d orbitals) 
%spin partial d densities of states for the 
%Mn(x) site respectively.}
}
\label{fig:unfav}
%\end{center}
\end{figure}

%%%%%%%%%%%%%%%%%%%%%%%%%%%%%%%%%%%%%%%%%%%%%%%%%%%%%%%%%%%%%%%%%%%%%%%%%%%%%%%%%
%%%%%%%%%%%%%%%%%%%%%%%%%%%%%%%%%%%%%%%%%%%%%%%%%%%%%%%%%%%%%%%%%%%%%%%%%%%%%%%%%
%%%%%%%%%%%%%%%%%%%%%%%%%%%%%%%%%%%%%%%%%%%%%%%%%%%%%%%%%%%%%%%%%%%%%%%%%%%%%%%%%

%%%%%%%%%%%%%%%%%%%%%%%%%%%%%%%%%%%%%%%%%%%%%%%%%%%%%%%%%%%%%%%%%%%%%%%%%%%%%%%%%%%% 
These results can be easily understood in terms of the OO reducing the overlap between the localized (SIC)
e$_g$ states and the 2p states of the adjacent oxygen atoms. 
This is reflected in the DOS of scenario 3 by the increased hybridization in the O 2p channel around
-0.8 Ry, over the whole width of the valence band and even in the
conduction band (Fig. \ref{fig:unfav}), in comparison with the DOS of scenario 1. 
%This is due to the fact that the full
%weight of the SIC localized e$_g$ states is shifted well below the valence band. 
Thus localized e$_g$ states which have lobes along the long manganese oxygen bonds will be energetically favourable 
configurations, while those that point along the short manganese oxygen bonds will correspond to 
energetically unfavourable configurations. 
Therefore the next two in the sequence of favourable scenarios are also antiferro OO 
involving in one case d$_{z^2-x^2}$ 
and d$_{y^2-z^2}$ (8.5 mRy above the ground state) and in the other case 
the d$_{3x^2-r^2}$ and d$_{3y^2-r^2}$ but with ferromagnetic spin arrangement (unfavourable by $\sim$ 1 mRy).
It is significant that even with the ferromagnetic spin arrangement the antiferro OO is lowest
in energy indicating again the importance of the JT effect. 
In summary, using the SIC-LSD theory it has been possible to investigate the orbital, spin and charge ordering 
of distorted LaMnO$_3$.
%The results obtained for the distorted structure are consistent with, and provide an {\it ab initio} understanding
%of, the highly intuitive picture of minimising the overlap between the localised e$_g$ states and the adjacent 
%oxygen p states \cite{Khomskii}.
%The ground state obtained is that of $Mn^{3+}$ antiferro orbitally ordered $d_{3x^2-r^2}$ and $d_{3y^2-r^2}$ 
%in the distorted A-type antiferromagnetic structure. The application of the SIC to cubic LaMnO$_3$ did not 
%induce orbital ordering as proposed in \cite{Khomskii}, which
%%and the degeneracy of the orbitals has not been lifted.
%leads us to conclude that structural distortions lead to OO in LaMnO$_3$. This we could further corroborate
%by the JT fit of our {\it ab initio} calculations and the finding that JT energies are larger than
%magnetic energies.
We find that our calculated values for the orbital energy depend strongly on the lattice distortion 
and are essentially independent of the magnetic order.  
We have used various numerical estimates to support our claim 
that the Jahn Teller interaction is the dominant effect 
in producing OO.  The lattice effects are big enough to account for 
the observed OO,  
meaning that it is not necessary to invoke additional contributions of an electronic origin.  
This, however, should be compared with the results of other papers,\cite{Medvedeva,Okamoto}  
where it is claimed that
OO could occur from electronic effects alone.  
We agree with this in so far that we also find an electronic effect of the correct symmetry.  
However, we find that the energy associated with the pure electronic effect is relatively small
%and that the size of the effect associated with the distortion agrees with independent experiments 
%and is large enough to give rise to the structural transition.
and that the size of the effect associated with the distortion is large enough to give rise 
to the structural transition. In addition, this could be corroborated by comparing with 
independent experiments. This is an important finding for the transport in the doped materials 
because as the electrons become delocalised the lattice is unable to respond fast enough and 
one reaches the large polaron regime. On the other hand OO that is purely electronic could 
coexist in the metallic phase giving rise to residual OO and extra contributions to the scattering.

\end{document}